\documentclass[prb,twocolumn,showpacs,amsmath,amssymb]{revtex4}
\usepackage{bm}
\usepackage[dvipdfmx]{graphicx}
\usepackage{color}

\begin{document}

\title{Engineering Dirac electrons emergent on the surface of a topological insulator}

\author{Yukinori Yoshimura$^1$}
\author{Koji Kobayashi$^2$}
\author{Tomi Ohtsuki$^2$}
\author{Ken-Ichiro Imura$^1$}

\affiliation{$^1$Department of Quantum Matter, AdSM, Hiroshima University, Higashi-Hiroshima, 739-8530, Japan}
\affiliation{$^2$Department of Physics, Sophia University, Chiyoda-ku, Tokyo, 102-8554, Japan}

\date{\today}

\begin{abstract}
The concept of topological insulator (TI)
has introduced a new point of view
to condensed-matter physics,
relating {\it a priori} unrelated subfields such as
quantum (spin, anomalous) Hall effects,
spin-orbit coupled materials,
some classes of nodal superconductors and superfluid $^3$He, etc.
From a technological point of view,
topological insulator is expected to serve as a platform for realizing
dissipationless transport in a non-superconducting context.
The topological insulator exhibits a gapless surface state
with a 
characteristic conic dispersion (a surface Dirac cone).
Here, we review peculiar finite-size effects
applicable to such surface states in TI nanostructures.
We highlight
the specific electronic properties of TI
nanowires and nanoparticles,
and in this context contrast
the cases of weak and strong TIs.
We study robustness of the surface and the bulk of TIs against disorder,
addressing the physics of Dirac and Weyl semimetals
as a new perspective of research in the field.
\end{abstract}


\maketitle

\section{Introduction}
Topological insulator
\cite{Moore,HasanKane,QIZhang}
enables dissipationless transport in a non-superconducting context.
It may allow for fixing ``neutrinos'' in semiconductor chips.
Known examples of the topological insulator
are such materials as
Bi$_2$Se$_3$, 
\cite{BiSe}
Bi$_2$Te$_3$, 
\cite{BiTe}
and many of their relatives.
\cite{Ando, AK_PRL}
These are three-dimensional (3D) bulk materials 
under the influence of relatively strong spin-orbit coupling.

In the bulk of a sample, 
a topological insulator is, at least superficially, not different from a normal band insulator, 
while on the surface, it is quite different.
A topological insulator exhibits a gapless surface state, 
the energy of which traverses the bulk energy gap. 
It is often said that the bulk energy gap of a topological insulator is ``inverted,'' 
\cite{KaneMele, BHZ}
in comparison with the normal one, 
though giving a precise meaning to this phrase 
needs further formulation of the bulk effective Hamiltonian.
\cite{Liu_nphys, Liu_PRB}
Whether the gap is inverted or not is 
specified by a winding (topological) number 
that encodes the bulk band structure, 
and is in one-to-one correspondence with 
whether the system exhibits a gapless surface state 
(bulk-boundary correspondence).
\cite{HG}

The surface state of a topological insulator exhibits a gapless spectrum 
often represented by the word, ``Dirac cone,''
described by an (effective)  2D gapless Dirac equation.
Dirac electrons represented by such a Dirac equation
must have an active spin degree of freedom, either real or fictitious.
In the case of graphene,
another platform for realizing such 2D Dirac electrons,
this role is played by the sublattice degrees of freedom. 
Here, on the surface of a topological insulator,
the spin is real, and since its direction is locked to the momentum
(spin-to-momentum locking),
the corresponding Dirac cone is often
said to be {\it helical}.
On generically curved surface of topological insulator samples of an arbitrary shape
the same spin is also locked to the local tangent to the surface
(spin-to-surface locking).
In topological insulator nanostructures
this second type of locking that is effective in real space
plays an important role in
determining the low-energy spectrum of the surface Dirac states.
In the first half of the paper
we review
such finite size effects associated with the spin-to-surface locking.

Another aspect of the 3D topological insulator is that
it has two subclasses:
weak and strong.
\cite{MooreBalents,FuKaneMele,Roy}
The strong topological insulator exhibits a single, or more generally,
an odd number of Dirac cones on its surface,
while its weak counterpart exhibits an even number of Dirac cones
in the surface Brillouin zone.
A single Dirac cone is robust against disorder, while an even number
of Dirac cones could be more easily destroyed.
\cite{Mong,Liu_physica}
However,
being robust does not necessarily mean {\it useful}.
\cite{dark, AT1}
Recall the case of semiconductor vs. metal.
Metal is always conducting, while semiconductor is either conducting or
insulating, depending on the concentration of impurities.
This fragility of the semiconductor makes it more useful than a metal, at least for some purposes,
say, for functioning a transistor.
The case of weak topological insulator is somewhat similar;
the fragility of the even number of Dirac cones makes {\it controllable}
the Dirac electrons emergent on the surface of a weak topological insulator.
In the second half of the paper
we argue that this controllability of the weak topological insulator
surface states
allows for constructing a dissipationless nanocircuit 
simply by patterning its surface
(by the use of lithography and etching).

\begin{figure}[htbp]
 \begin{tabular}{c}
  \put(-10,110){(a)}
  \includegraphics[width=70mm, bb=0 0 259 165]{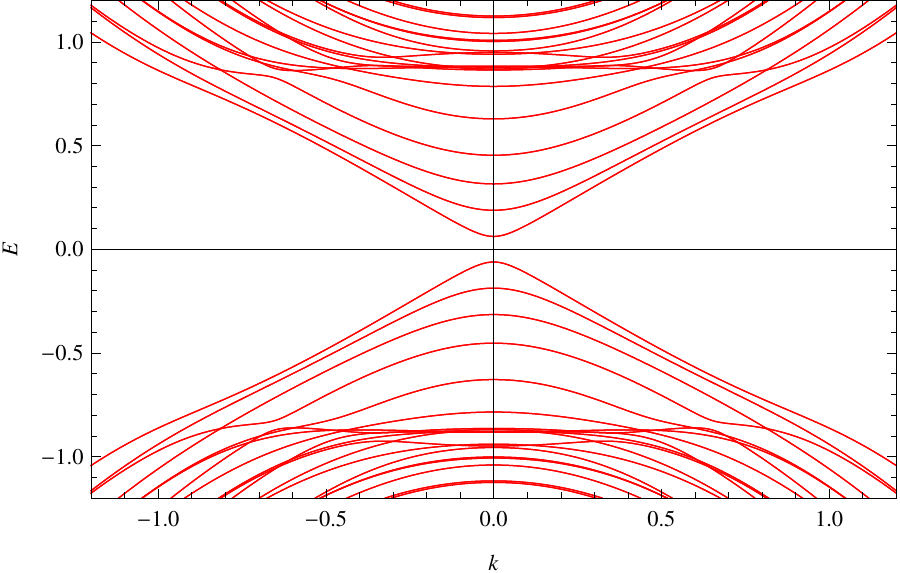}
  \\
  \put(-10,110){(b)}
  \includegraphics[width=70mm, bb=0 0 259 166]{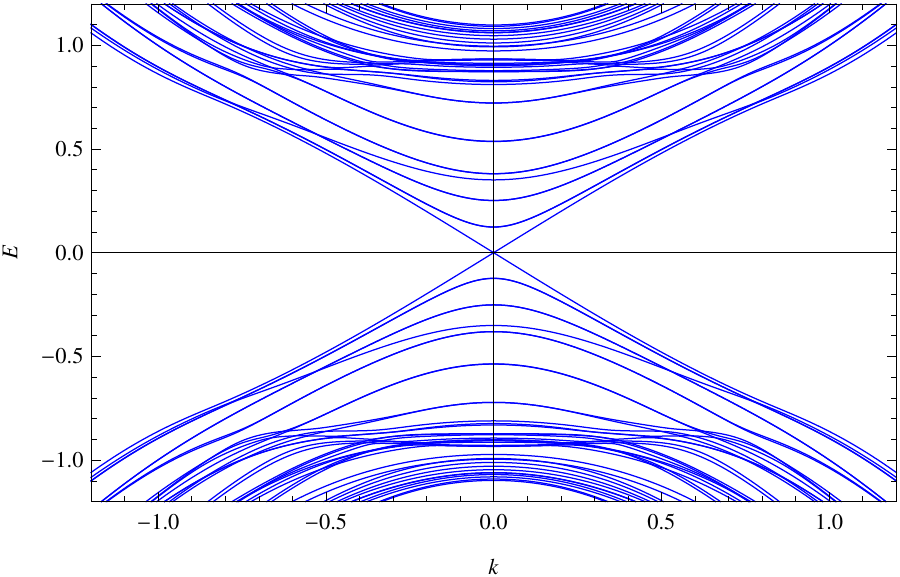}
 \end{tabular}
\vspace{-2mm}
\caption{Schematic energy spectrum of a prism-shaped 
topological insulator nanowire.
(a) As a consequence of the spin Berry phase $\pi$,
the spectrum of the surface state is gapped.
(b) the same spectrum becomes gapless in the presence of an external flux
$\pi$ inserted along the axis of the prism 
so as to cancel the Berry phase.}
\label{prism}
\end{figure}

\begin{figure*}[htbp]
 \begin{tabular}{c@{\hspace{15mm}}c}
  \put(-20,110){(a)}
  \includegraphics[width=40mm, bb=0 0 1304 1311]{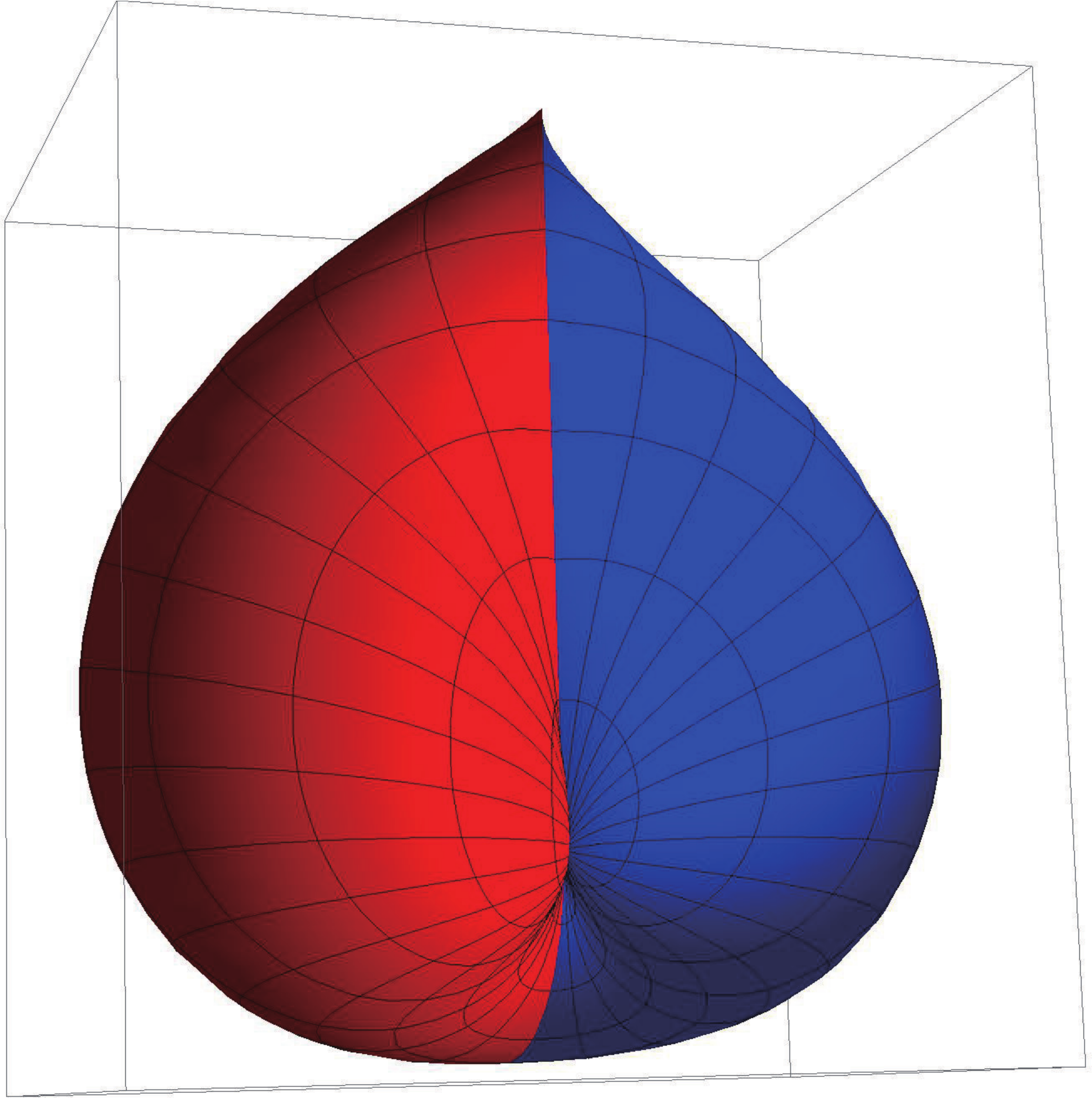}  &
  \put(-20,110){(b)}
  \includegraphics[width=60mm, bb=0 0 1214 884]{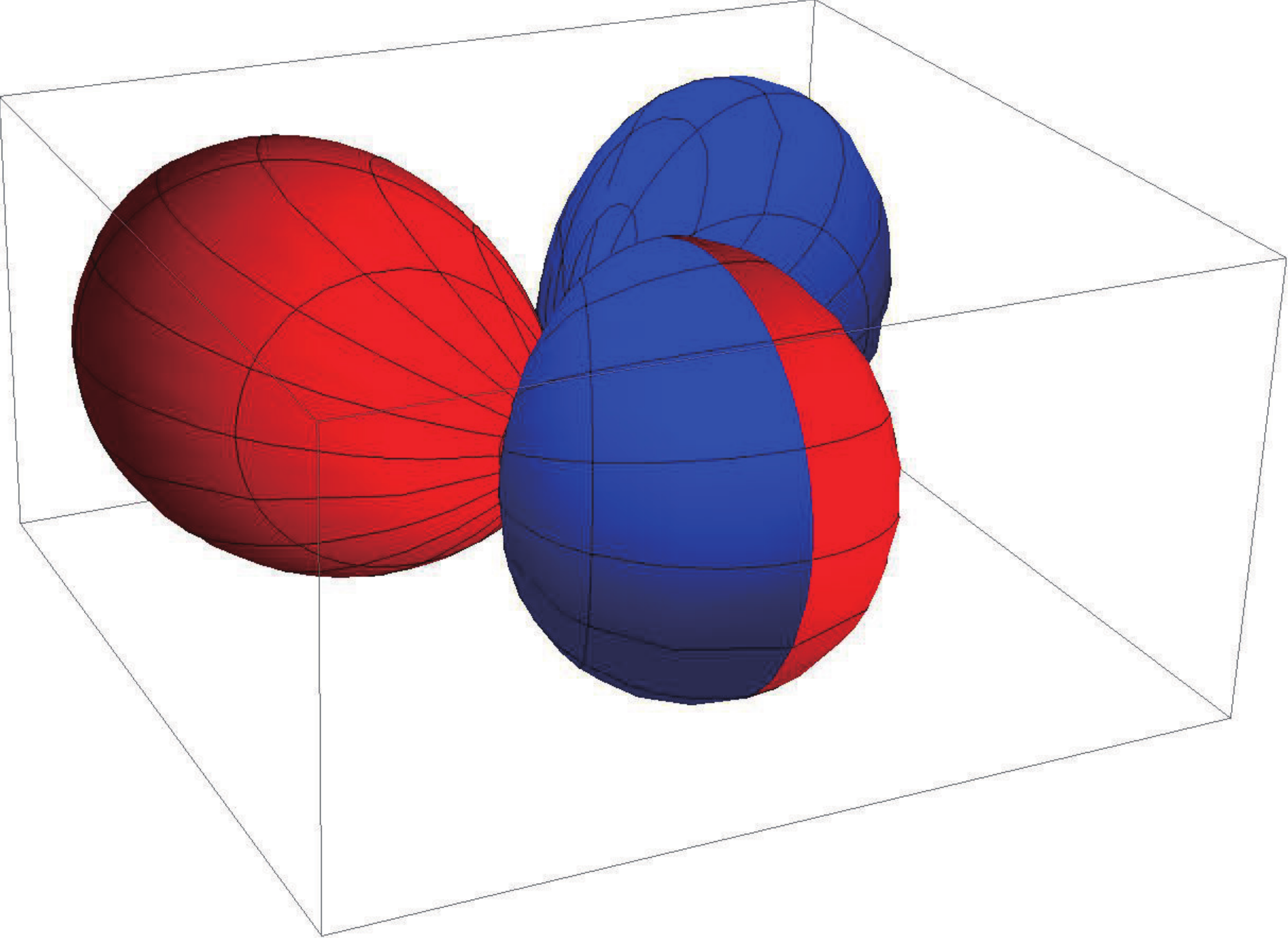}
 \end{tabular}
\vspace{-2mm}
\caption{Topological insulator nanoparticles as an ``artificial atom.''
Spatial profile of some low-lying 
electron states of this artificial atom.
The antiperiodic version of the (a) $s$-type, (b) $p$-type orbitals are shown.
To highlight their characters, the orbitals are painted in red (in blue)
when the real part of the wave function is positive (negative).
}
\label{atom}
\end{figure*}

\section{Peculiar finite size effects in topological insulator surface states ---
part 1:  the strong case}

The helical Dirac electron emergent on the surface of a topological insulator
is described,
in the $\bm k\cdot\bm p$ approximation,
by the following effective
surface Dirac Hamiltonian,
\begin{align}
H_{\rm surf} = v (p_y \sigma_z - p_z \sigma_y).
\label{H_surf_flat}
\end{align}
Here, the surface is chosen to be normal to the $x$-axis,
and $v$ is the velocity of a Dirac electron.
The $2\times 2$ matrix structure of Eq. (\ref{H_surf_flat})
is due to the real spin degree of freedom.
The form of Eq. (\ref{H_surf_flat})
implies that momentum eigenstates have a spin
oriented perpendicular to the direction of the momentum.
Or, the other way around
the spin is locked to the momentum: spin-to-momentum locking.
This is the reason why the corresponding Dirac cone is said to be {\it helical}.

Another feature
encoded in the explicit form of Eq. (\ref{H_surf_flat})
is that the spin of the momentum eigenstates
has no out-of-plane component.
The spin is indeed locked (in-plane) to the surface.
This feature, sometimes represented by the word,
spin-to-surface locking
holds also true
on generically curved surfaces of a sample of arbitrary shape.
\cite{Ashvin_PRB, unified}
Here, we will focus on cases of 
{\it cylindrical} and {\it spherical} shaped samples, in particular.

\subsection{Intrinsic Aharonov-Bohm effect}

Here, let us first focus on the {\it cylindrical} case.
By that we mean targeting physically a topological insulator nanowire.
On the cylindrical surface of such a nanowire
Eq. (\ref{H_surf_flat}) is modified as,
\cite{AT2}
\begin{align}
H_{\rm surf} = v
\left[
\begin{array}{cc}
0 & -i p_z + {\hbar\over R} \left(-i{\partial \over \partial\phi}+{1\over 2}\right) 
\\ 
i p_z + {\hbar\over R} \left(-i{\partial \over \partial\phi}+{1\over 2}\right) & 0
\end{array}
\right],
\label{H_surf_cyl}
\end{align}
where $R$ is the radius of the cylinder.
The additive factor $1/2$ that appears in the two off-diagonals
of Eq. (\ref{H_surf_cyl})
is a spin Berry phase encoding the constraint on the direction of spin
as a consequence of the spin-to-surface locking.
When a Dirac electron moves around the curved, cylindrical surface of
the topological insulator nanowire,
the spin-to-surface locking constrains
the direction of the spin
to follow the change of the tangential surface.
As a result,
when an electron completes a $2\pi$ rotation in the configuration space,
its spin also performs a $2\pi$ rotation in the spin space,
while the spin is naturally double-valued.
This explains the origin of the additive factor $1/2$,
or in other words, a Berry phase $\pi$.
\cite{Ashvin_PRB, Mirlin, AT1, AT2}

The orbital part of an 
electron state on the surface of a cylindrical
topological insulator is represented as
\begin{align}
 \psi (\phi,z)= e^{{i \over \hbar}(L_z\phi + p_z z)},
\end{align}
where
$L_z =p_\phi R$ is the orbital angular momentum in the $z$-direction,
along the axis of the cylinder.
The Berry phase $\pi$
modifies boundary condition
with respect to the polar angle $\phi$ from {\it periodic} 
to {\it anti-periodic}, {\it i.e.},
\begin{align}
 e^{{i \over \hbar} L_z (\phi+2\pi)}\times (-1)= e^{{i \over \hbar} L_z \phi}.
\end{align}
This leads to the following quantization rule
of the orbital angular momentum:
\begin{align}
 {L_z \over \hbar}=\pm {1\over 2}, \pm {3\over 2}, \cdots.
\label{half_odd}
\end{align}
In this half-integral quantization $L_z=0$,
and therefore, $p_\phi =0$ is not allowed.
On the other hand,
the spectrum of the surface states takes the following
Dirac form,
\begin{align}
 E_{\rm surf} = \pm v\sqrt{p_\phi^2+p_z^2}.
\label{E_surf}
\end{align}
Combining Eqs. (\ref{half_odd}) and (\ref{E_surf}),
one is led to convince oneself that
the spectrum of the cylindrical topological insulator is generically {\it gapped}
[see also a gapped spectrum shown in FIG. 1 (a)].
The spectra shown in FIG. 1 are
calculated using tight-binding implementation of the bulk 3D 
topological insulator of a rectangular prism shape.
\cite{AT2}

The magnitude of this energy gap does not decrease exponentially
as the increase of the size of the system,
in sharp contrast to the case of the usual hybridization gap.
\cite{Lu10}
The energy gap $\Delta E = v \hbar/R \propto R^{-1}$
decays only algebraically as a function of the size of the system.
In typical TIs (Bi$_2$Se$_3$, Bi$_2$Te$_3$), the Fermi velocity $v \simeq 4\times 10^5{\rm m/s} \simeq c/7500$,
and the gap is expected to be visible for $R \lesssim 10^{-7}$m, at 1 K.

The spin Berry phase $\pi$ can be interpreted
such that
a fictitious magnetic flux is induced and pierces the cylinder.
Indeed,
the additive factor $1/2$ that appears in the two off-diagonals
of Eq. (\ref{H_surf_cyl})
can be regarded as a vector potential
\begin{align}
\bm A = {\Phi \over r}\hat{\bm \phi},
\end{align}
created around a flux tube of strength $\Phi$,
identical to half of a flux quantum $\Phi_0=h/e$.
This corresponds to the Berry phase,
\begin{align}
\alpha = 2\pi {\Phi\over\Phi_0} = \pi.
\label{Berry}
\end{align}

An electron on the cylindrical surface does not touch the fictitious flux itself,
since the fictitious flux tube is at the deep inside of the cylinder.
Yet, quantum mechanically,
the spectrum of the surface state is still influenced by this effective flux;
the electron feels a vector potential, and its spectrum affected by the vector potential,
even when there is no magnetic field 
at any positions where the electron is allowed to exist
(Aharonov-Bohm effect).
Here, the effective flux is induced by a constraint on spin of the surface state.
Thus,
the electron on the cylindrical surface induces an effective flux, while
its quantum mechanical motion is influenced by the flux the electron itself created;
therefore, the name {\it intrinsic} Aharonov-Bohm effect.
To confirm this scenario
we repeated the same numerical simulation
in the presence of a real external magnetic flux
designed to cancel the fictitious one; Eq. (\ref{Berry}).
As expected,
the spectrum of the surface state becomes gapless,
once the Berry phase $\pi$ is cancelled by the external flux
[see FIG. 1 (b)].

Related experimental results in Bi$_2$Se$_3$ nanowire
have been reported in Refs.~\onlinecite{TI_nanowire1},~\onlinecite{TI_nanowire2}.

\subsection{Topological insulator nanoparticles as artificial atom}

Magnetic monopoles do not exist in nature. 
This is what we learn in courses of elementary electromagnetism, 
and the statement is, of course, true at the microscopic level.
In a matter, however, Maxwell equations are modified,
or at least it is convenient to replace them with effective equations.
Then, depending on the nature of effective media,
analogues of a magnetic monopole can appear.
This indeed happens in case of the spherical topological insulator.

Similarly to the cylindrical case [Eq. (\ref{H_surf_cyl})],
on a spherical surface of a topological insulator
Eq. (\ref{H_surf_flat}) is modified to
\cite{spherical}
\begin{align}
H_{\rm surf} = {v\over R}
\left[
\begin{array}{cc}
0  & -\partial_\theta + {i\partial_\phi\over \sin\theta} - {1\over 2} \cot {\theta \over 2} \\
\partial_\theta + {i\partial_\phi\over \sin\theta} - {1\over 2} \tan {\theta \over 2} & 0
\end{array}
\right].
\label{H_surf_sphe}
\end{align}
There are corrections due to the spin Berry phase in the off diagonals of
Eq. (\ref{H_surf_sphe}).
Here, in the spherical case,
they are understood as a vector potential
associated with a magnetic monopole.
The explicit form of the vector potential differs, 
depending on the choice of the gauge such that
\begin{align}
\bm A_{\rm I} = {g \over 4\pi r} \tan {\theta \over 2}\ \hat{\bm \phi}
\end{align}
when the singularity (the Dirac string) is chosen on the $-z$ axis,
while
\begin{align}
\bm A_{\rm II} =  {-g \over 4\pi r} \cot {\theta \over 2}\ \hat{\bm \phi}
\end{align}
when the Dirac string is on the $+z$ axis.
In consistent with the Dirac quantization condition,
Eq. (\ref{H_surf_sphe})
corresponds to the case of an effective magnetic monopole
of strength $g=\pm 2\pi$
(the smallest value compatible with the Dirac quantization condition).
The electron on the surface of a {\it spherical} topological insulator
behaves as if there is an effective {\it magnetic monopole} at the center of the sphere.

The spherical topological insulator
models naturally a topological insulator nanoparticle.
Taking it, therefore, as an artificial atom, 
let us focus on its low-lying electronic levels.
The angular part of the wave function is described by
an anti-periodic analogue of the spherical harmonics.
A few examples are shown in FIG. \ref{atom},
where
$\bm \alpha (\theta,\phi) = e^{i m\phi} \bm \alpha_{nm} (\theta)$
represents the angular dependence of a surface eigenfunction,
specified by quantum numbers $n$, $m$.
An analogue of the $s$-orbital is
\begin{align}
\bm \alpha_{0{1\over 2}} =
\left[
\begin{array}{l}
\alpha_{0{1\over 2}+} \\
\alpha_{0{1\over 2}-}
\end{array}
\right]
={1\over\sqrt{4\pi}}
\left[
\begin{array}{r}
\cos {\theta\over 2} \\[2mm]
-\sin {\theta\over 2}
\end{array}
\right],
\end{align}
[see FIG. \ref{atom}, panel (a)],
while
\begin{align}
\bm \alpha_{0{3\over 2}} =
\sqrt{3\over 2\pi}
\left[
\begin{array}{r}
\sin {\theta\over 2}\ \cos^2\!{\theta\over 2} \\[2mm]
- \sin^2\!{\theta\over 2}\ \cos {\theta\over 2}
\end{array}
\right]
\end{align}
can be regarded as an anti-periodic version of the
$p$-orbital [panel (b)].
Further details on such ``monopole harmonics''
\cite{monop}
and the spectrum of the artificial atom are given in Ref. \onlinecite{spherical}.

\begin{figure}[htbp]
\includegraphics[width=70mm, bb=0 0 612 389]{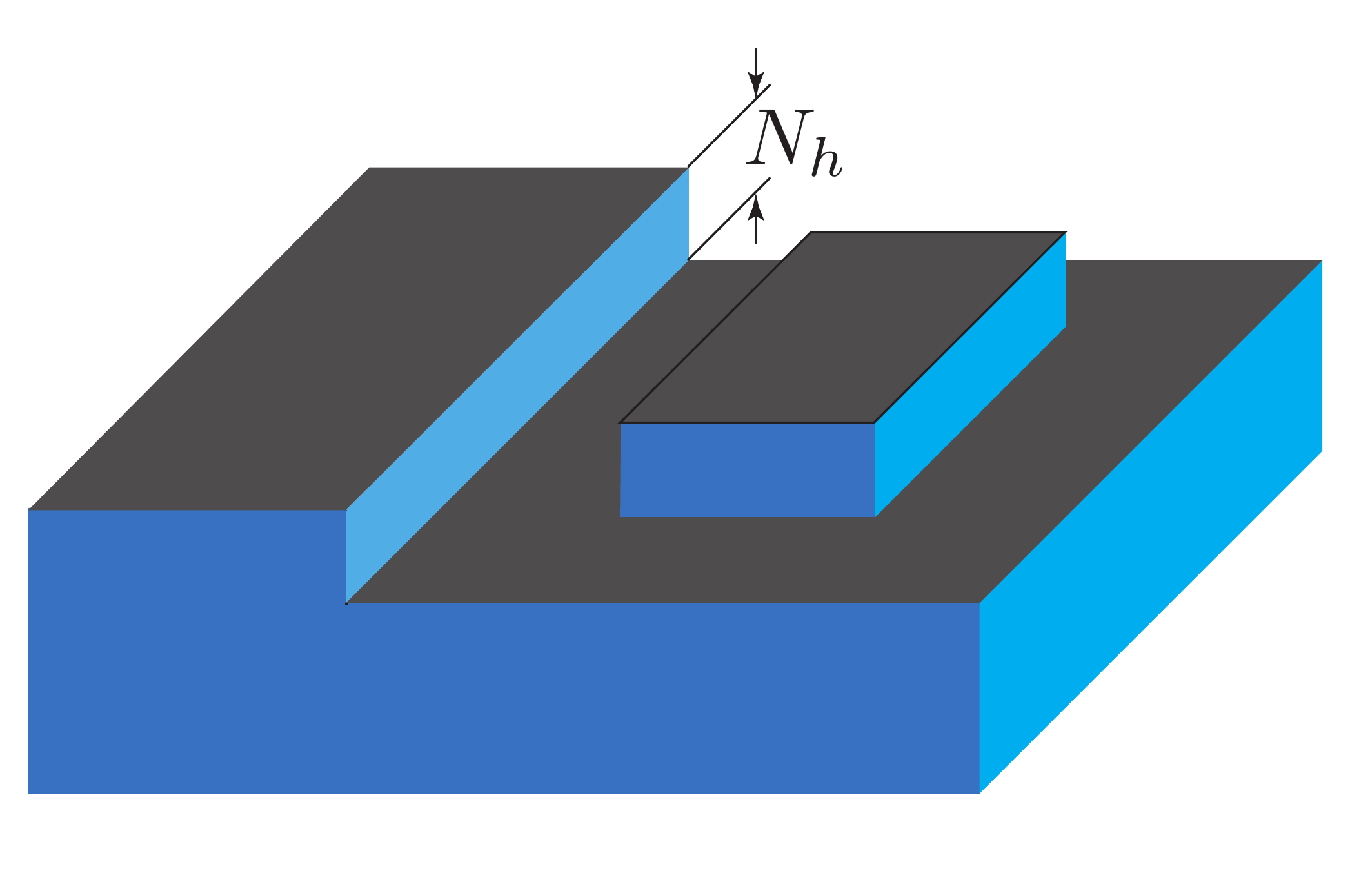}
\vspace{-2mm}
\caption{Pattering the surface of a weak topological insulator
allows for constructing nanocircuits of 1D dissipationless channels.
See also Ref. \onlinecite{dark}.
}
\label{schema}
\end{figure}

\begin{figure}[htbp]
 \begin{tabular}{c}
 \put(-20,90){(a)}
  \includegraphics[width=70mm, bb=0 0 384 259]{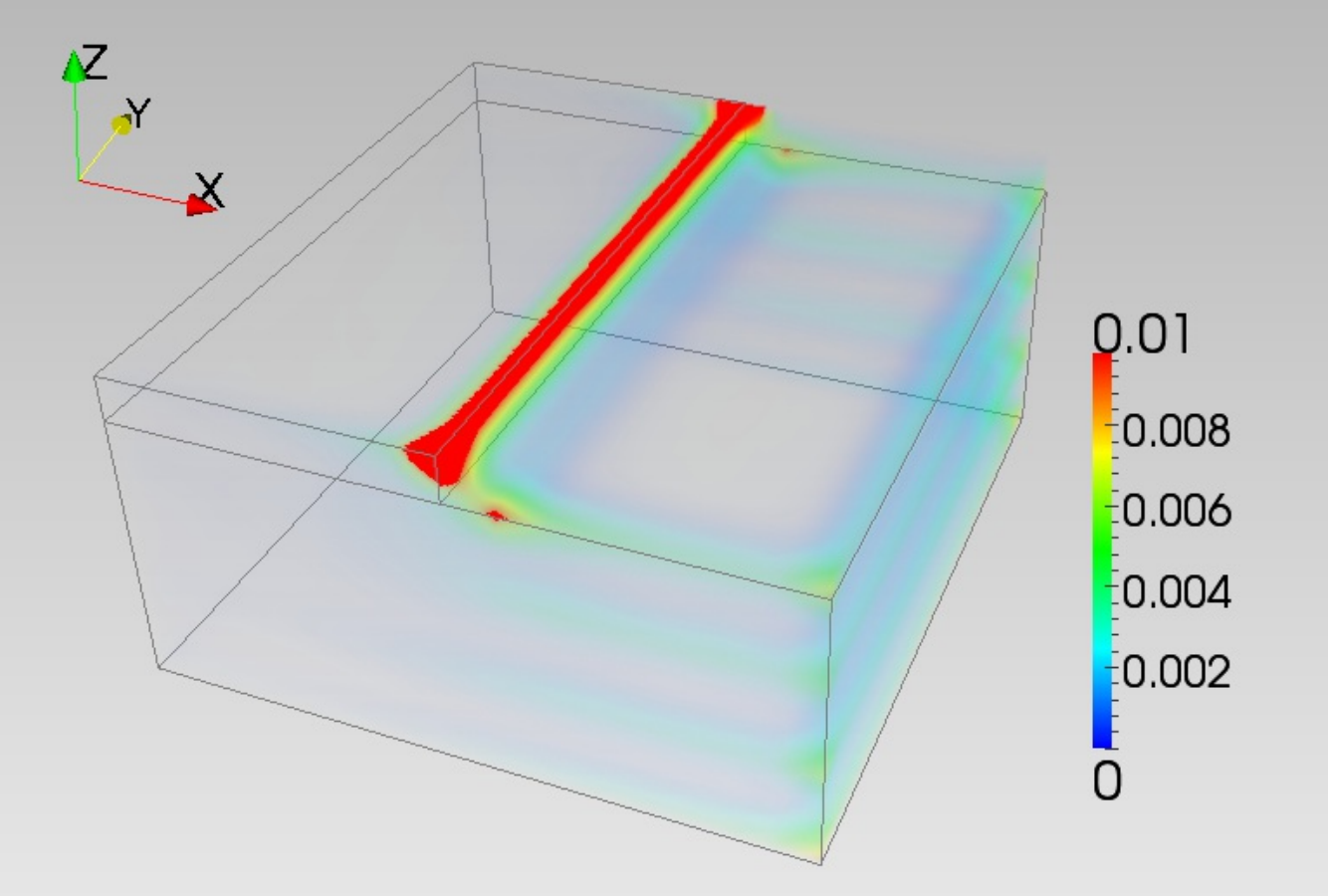}
  \\
   \put(-20,90){(b)}
  \includegraphics[width=70mm, bb=0 0 388 257]{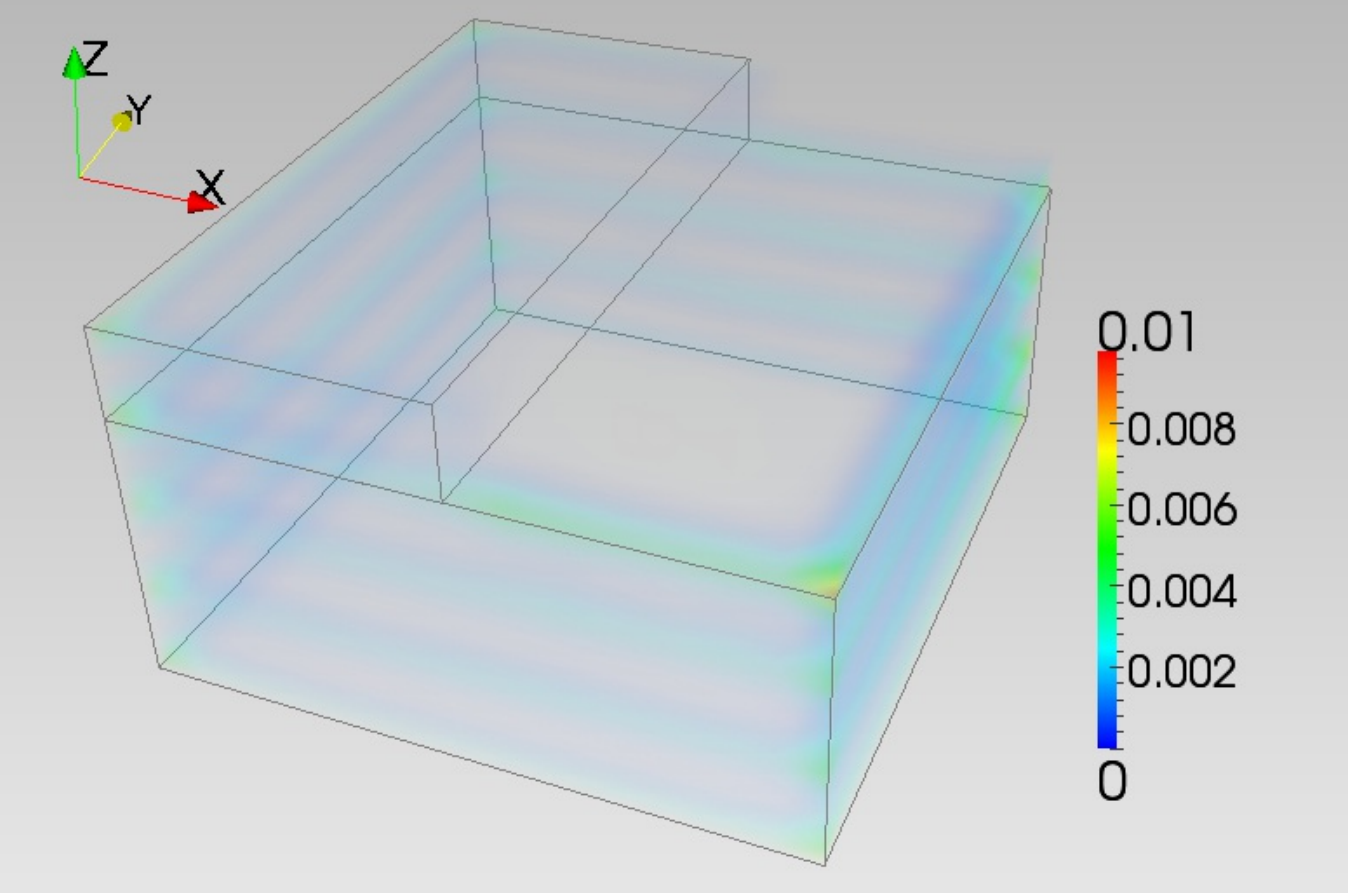}
 \end{tabular}
\vspace{-2mm}
\caption{Spatial profile of low-lying electronic wave function
along a step of height (a) one, (b) two.
(a) realizes a perfectly conducting channel.
See also Ref. \onlinecite{dark}.
}
\label{step}
\end{figure}

\section{Peculiar finite size effects in topological insulator surface states ---
part 2: the weak case}

The specificity of a 
3D topological insulator is that it can be either weak or strong
as mentioned in the Introduction.
The strong topological insulator (STI) exhibits typically a single Dirac cone, 
while the weak topological insulator (WTI) 
exhibits an even number of Dirac cones
in the surface Brillouin zone.
A single Dirac cone cannot be confined (Klein tunneling),
while an even number of Dirac cones
can be confined.
Therefore, in a STI
the single Dirac cone is extended over all the surfaces.
That is actually the situation we have considered so far in Sec. II.
In a WTI, on the contrary, some surfaces are metallic
(gapless as a consequence of an even number of Dirac cones), 
while others are not (no Dirac cone on such surfaces).

The WTI is characterized by three weak indices $\nu_1$, $\nu_2$, $\nu_3$.
\cite{FuKane07}
The surfaces normal to the direction specified by 
$\bm\nu = (\nu_1, \nu_2, \nu_3)$
are gapped (no Dirac cone).
In this sense,
the WTI is by its nature {\it anisotropic},
and can be regarded as a {\it layered material}.
In the following we demonstrate that
the WTI surface states are susceptible to
various even/odd features
in regard to the number of atomic layers
stacked in the direction of $\bm\nu$.

\subsection{Even/odd feature}

To highlight the even/odd feature
we first consider the spectrum of a WTI sample with
top, bottom and side surfaces.
The top and bottom surfaces are oriented normal to the $z$-axis.
To realize a typical situation
we assume that the top and bottom surfaces are {\it gapped} surfaces (no Dirac cone).
We then consider 
electronic states on a side surface
here chosen to be on the $zx$-plane.
The two Dirac cones are typically located at $k_z=0$ and at $k_z=\pi$.
Low-energy electron states at or in the vicinity of the Dirac points 
may be represented by a plane wave:
\begin{align}
\psi_1 &= e^{i(k_x x+q_1 z)},
\nonumber \\
\psi_2 &= e^{i[k_x x+(\pi+q_2)z]}.
\end{align}
Here, the crystal momentum $q_1$ or $q_2$
of the electron is measured from the corresponding Dirac cone.
Now, if we consider the presence of top and bottom surfaces,
located, say, respectively at $z=N_z$ and $z=1$,
we need to confine the above electron in the region of $1\le z\le N_z$,
{\it i.e.},
we impose the boundary condition 
such that the wave function of the surface state
$\psi$ that may be expressed as a linear combination of 
$\psi_1$ and $\psi_2$ satisfies
\begin{align}
\psi (z) &= c_1 \psi_1 + c_2 \psi_2,
\nonumber \\
\psi (z=0) &= 0,\ 
\psi (z=N_z +1) = 0. 
\end{align}
To cope with the boundary condition at $z=0$, we choose the constants
$c_1$ and $c_2$ such that $c_2= - c_1=1$.
Also, at a given energy $E$
one can set $q_1$ and $q_2$ such that
$q_2 = -q_1\equiv q$.
We are then left with
\begin{align}
\psi (z) = e^{ik_x x} \left[ e^{iqz} - (-1)^{z}e^{-iqz} \right],
\label{psi_z}
\end{align}
and this must vanish at $z=N_z +1$. Therefore,
for $N_z$ odd,
\begin{align}
q=\pm {\pi\over 2(N_z+1)}2n
\label{odd}
\end{align}
where
$n=0, \pm 1, \pm 2, \cdots$
can be an arbitrary integer.
Similarly, for $N_z$ even,
the vanishing of Eq. (\ref{psi_z}) at $z=N_z+1$ implies
\begin{align}
q=\pm {\pi\over 2(N_z+1)}(2m-1)
\label{even}
\end{align}
where $m=1, 2, 3, \cdots$, and $\pm (2m-1)$ is an arbitrary odd integer.
Since the spectrum of the surface state is given as
\begin{align}
E=\pm v\sqrt{k_x^2+q^2},
\end{align}
Eqs. (\ref{odd}), (\ref{even})  
signify that the surface spectrum is gapless for $N_z$ odd,
while it is gapped when $N_z$ is even.
\cite{mayuko1, arita}
Also, note that
replacing $N_z$ with $N_h$,
one can equally apply
Eqs. (\ref{odd}) and (\ref{even})
for characterizing the 1D helical modes that appear
along a step formed on the surface of a WTI
\cite{dark};
{\it c.f.} FIG. 3 and discussion given in the next section.

\subsection{Dissipationless nanocircuit, or 1D perfectly conducting channel}

We have so far argued that WTIs are susceptible to a specific type of size effect 
that is strongly dependent on the parity of the number of layers
stacked in the direction of $\bm\nu = (\nu_1, \nu_2, \nu_3)$.
Here, we demonstrate that
this even/odd feature makes indeed a WTI more useful than an STI.
 We consider the following Wilson-Dirac-type effective Hamiltonian on the cubic lattice,
\begin{align}
   H_{\rm bulk}&= v \sum_{\mu=x,y,z} \sin k_{\mu}\  \tau_x \otimes \sigma_{\mu} +  m({\bm k})\ \tau_z \otimes 1_2,  \nonumber\\
   m({\bm k}) &= m_0 + \sum_{\mu=x,y,z} 2 m_{2\mu} (1 - \cos k_{\mu}),
\end{align}
where two types of Pauli matrices $\sigma$ and $\tau$ represent real and orbital spins, respectively, and $1_2$ is $2\times 2$ identity matrix.
 We have chosen the parameters $m_0 = -v$, $m_{2x} = m_{2y} = 0.5v$, $m_{2z} = 0.1v$, and $v = 2$, 
so that a WTI with weak indices,
$\bm\nu = (0,0,1)$
is realized.
\cite{mayuko1}
 By simply patterning the surface of a WTI
(see FIG. \ref{schema})
one can realize a nanocircuit of 1D perfectly conducting channel.

The even/odd feature
on the spectrum of a WTI samples with side surfaces is
equally applied to 
a system with atomic scale steps on the gapped surfaces.
\cite{dark}
In FIG. \ref{step}
we show the spatial profile of the wave function of some low-lying states
in a geometry with an atomic scale step.
In panel (a) the height of the step $N_h$ is 1, while in panel (b)
the step is 2 atomic scale high.
As shown in the two panels,
the wave function has strong amplitude along the step when $N_h$ is odd,
while visibly no weight in the step region
when $N_h$ is even.

These 1D channels that appear when $N_h$ is odd, is also shown to be
helical, and robust against non-magnetic disorder.
\cite{takane,dark,Stern}
In a supplemental material\cite{suppl}
we demonstrate the robustness of such an odd number of channels
against disorder.
Related to this, the robustness of the underlying WTI has been also
studied;
\cite{KOI, Bjoern}
see also Sec. IV.
In any case such a nanocircuit that appears on a patterned surface
of WTI is perfectly conducting not only in the clean limit
but also in the presence of disorder.
 Therefore the 1D perfectly conducting channel is stable at least within the phase coherence length;
to estimate the order of its magnitude,
note that this is
typically $200$ nm, for Bi$_2$Se$_3$ at $5$ K.\cite{Matsuo}

Recently, WTI has been discovered experimentally in a bismuth-based
layered bulk material, Bi$_{14}$Rh$_3$I$_9$,
built from grapheme-like Bi-Rh sheets.
\cite{WTI_exp}
Other candidates for WTIs are layered semiconductors,
\cite{Felser}
superlattices of topological and normal insulators.
\cite{FIH, WTI_super1, WTI_super2}
The superlattice has been considered also in different contexts;
{\it e.g.}, as a protocol for realizing a Weyl semimetal,
\cite{Burkov}
and is indeed fabricated experimentally.
\cite{super_exp}
Still another possibility to realize a WTI-like situation is to use
topological crystal insulators.
\cite{Fu_TCI, TCI_Nature1, TCI_Nature2, TCI_Nature3}
These are close relatives of the standard topological insulator, protected by
crystalline symmetries instead of the time reversal symmetry.


\section{Concluding remarks}

In Sec. II
we have highlighted the {\it enhanced} size effect 
characteristic to Dirac electrons emergent on the surface of topological insulators.
In the classification of 3D topological insulators
to weak and strong categories,
this size effect was particularly applicable to the strong case.
This type of size effect also plays the role of protecting the surface state
from surface roughness. \cite{noninvasive}
In Sec. III we discussed that
weak topological insulators are susceptible to a different type of
size effect
that shows an even/odd feature with respect to the number of layers
in the stacking direction.
This even/odd feature indeed makes {\it weak} topological insulators more {\it useful} than
{\it strong} topological insulators.
By simply patterning the surface of a weak topological insulator,
one can achieve dissipationless transport in a non-superconducting context,
{\it i.e.},
nanocircuits of 1D perfectly conducting channels.

While in this paper we have focused on the clean limit,
the helical nature of the surface Dirac electron
is responsible for its robustness against disorder.	
Because of the spin-to-momentum locking, backscattering is forbidden;
for an incident state with momentum $\bm k$,
the reflected state with $-\bm k$ state has a spin part orthogonal to that of the initial state,
hence absence of backscattering.
\cite{ANR}
The gapless 2D Dirac semimetal is known to possess
unusual robustness against disorder.
\cite{KN,Bardarson,Tworzydlo08}
A slightly different question is 
how robust is the topological classification of bulk
in the clean limit against disorder.
In Ref.~\onlinecite{KOI} we have addressed this issue numerically
and show that
the concept of weak as well as strong topological insulators
remains valid at finite disorder.
Then, in a more recent paper, Ref.~\onlinecite{DSM},
we have extended this study to show that
not only 
such distinct topological phases but also a 3D Dirac semimetal that appears
at their phase boundary
show some robustness against disorder.
A similar result has been obtained also for a Weyl semimetal.
\cite{Ominato}
To highlight
such Dirac and related Weyl semimetals, especially in the context of their
robustness against disorder
is a new trend in the field
both theoretically 
\cite{DSM_theo1, DSM_theo2}
and experimentally.
\cite{DSM_exp1, DSM_exp2}

\acknowledgments
The authors acknowledge fruitful scientific interactions with Yositake Takane, Akihiro Tanaka,
Takahiro Fukui and Igor Herbut. One of the authors (T. O.) has been supported by 
Grants-in-Aid for Scientific Research (C) (Grant No. 23540376) and Grants-in-Aid No. 24000013.

\bibliography{STAM5}

\end{document}